\documentclass[10pt]{article}

\usepackage{amsthm,amsmath,latexsym,color,subfigure,setspace, multirow,soul}
\usepackage{url}\RequirePackage[colorlinks,citecolor=black, linkcolor=black,urlcolor = black]{hyperref}
\usepackage{amssymb}

\usepackage{graphicx}
\usepackage{epstopdf}
\usepackage{pdflscape}

\usepackage{cite}
\usepackage{enumerate}

\usepackage{xr}

\usepackage[labelfont=bf,labelsep=period,justification=raggedright]{caption}

\makeatletter
\renewcommand{\@biblabel}[1]{\quad#1.}
\makeatother

\usepackage{adjustbox}
\usepackage{array}
\usepackage{relsize}
\usepackage{sidecap}
\usepackage{bm}
\usepackage{bbm}
\usepackage{multicol}
\usepackage{float}
\usepackage{booktabs}
\usepackage{multirow,multicol}

\usepackage{tikz}
\usetikzlibrary{patterns}

\newtheorem{theorem}{Theorem}
\newtheorem*{theorem*}{Theorem}
\newtheorem{corollary}{Corollary}

\newtheorem*{claim*}{Claim}
\theoremstyle{definition}
\newtheorem{definition}[theorem]{Definition}
\theoremstyle{AppDefinition}
\newtheorem{AppDefinition}{Definition}[]
\theoremstyle{remark}

\def\eq#1{(\ref{#1})}

\def\beginmat{ \left( \begin{array} }
\def\endmat{ \end{array} \right) }

\def\cond{\, | \,}
\newcommand*\diff{\mathop{}\!\mathrm{d}}
\newcolumntype{P}[1]{>{\centering\arraybackslash}p{#1}}

\date{}

\newcommand{\bv}{\mathbf{v}}

\newcommand{\by}{\mathbf{y}}

\newcommand{\bu}{\mathbf{u}}

\newcommand{\bF}{\mathbf{F}}

\newcommand{\bI}{\mathbf{I}}

\newcommand{\T}{\intercal}

\newcommand{\wh}{\widehat}

\newcommand{\cN}{\mathcal{N}}

\newcommand{\bvarepsilon}{\boldsymbol\varepsilon}

\newcommand{\balpha}{\boldsymbol\alpha}
\newcommand{\bmu}{\boldsymbol\mu}

\newcommand{\bSigma}{\boldsymbol\Sigma}

\newcolumntype{R}[2]{%
    >{\adjustbox{angle=#1,lap=\width-(#2)}\bgroup}%
    l%
    <{\egroup}%
}

\allowdisplaybreaks

\begin{document}

\begin{flushleft}
{\Large
\textbf{Predicting Clinical Outcomes in Glioblastoma:\\ An Application of Topological and Functional Data Analysis}
}
\newline
\\
Lorin Crawford\textsuperscript{1-3,$\dagger$}, Anthea Monod\textsuperscript{4,$\dagger$}, Andrew X.~Chen\textsuperscript{5}, Sayan Mukherjee\textsuperscript{6-9}, and Ra\'ul Rabad\'an\textsuperscript{5}
\\
\bigskip
\bf{1} Department of Biostatistics, Brown University, Providence, RI, USA
\\
\bf{2} Center for Statistical Sciences, Brown University, Providence, RI, USA
\\
\bf{3} Center for Computational Molecular Biology, Brown University, Providence, RI, USA
\\
\bf{4} Department of Applied Mathematics, Tel Aviv University, Tel Aviv, Israel
\\
\bf{5} Department of Systems Biology, Columbia University, New York, NY, USA 
\\
\bf{6} Department of Statistical Science, Duke University, Durham, NC, USA
\\
\bf{7} Department of Computer Science, Duke University, Durham, NC, USA
\\
\bf{8} Department of Mathematics, Duke University, Durham, NC, USA
\\
\bf{9} Department of Bioinformatics \& Biostatistics, Duke University, Durham, NC, USA 
\\
\bigskip
$\dagger$ Corresponding E-mail: lorin\_crawford@brown.edu; antheam@tauex.tau.ac.il
\end{flushleft}


\section*{Abstract}

Glioblastoma multiforme (GBM) is an aggressive form of human brain cancer that is under active study in the field of cancer biology.  Its rapid progression and the relative time cost of obtaining molecular data make other readily-available forms of data, such as images, an important resource for actionable measures in patients.  Our goal is to utilize information given by medical images taken from GBM patients in statistical settings.  To do this, we design a novel statistic---the smooth Euler characteristic transform (SECT)---that quantifies magnetic resonance images (MRIs) of tumors. Due to its well-defined inner product structure, the SECT can be used in a wider range of functional and nonparametric modeling approaches than other previously proposed topological summary statistics. When applied to a cohort of GBM patients, we find that the SECT is a better predictor of clinical outcomes than both existing tumor shape quantifications and common molecular assays.  Specifically, we demonstrate that SECT features alone explain more of the variance in GBM patient survival than gene expression, volumetric features, and morphometric features.  The main takeaways from our findings are thus twofold. First, they suggest that images contain valuable information that can play an important role in clinical prognosis and other medical decisions. Second, they show that the SECT is a viable tool for the broader study of medical imaging informatics.



\section{Introduction}\label{intro}

The field of radiomics is focused on the extraction of quantitative features from medical magnetic resonance images (MRIs), typically constructed by tomography and digitally stored as shapes or surfaces.  Quantifying geometric features from shapes in a way that is amenable to computational analyses has been a long-standing and fundamental challenge in both statistics and radiomics.  Overcoming such a challenge would provide significant breakthroughs in broader scientific disciplines with the potential for real, practical impact.  One particularly important application, where a viable quantification of shapes is needed, is the study of glioblastoma multiforme (GBM)---a glioma that materializes into aggressive, cancerous tumor growths within the human brain.  GBM is a disease that is currently under active research in oncology; it is marked by characteristics that are not common in other cancers, such as spatial diffusivity and molecular heterogeneity.  In human patients, it is a rapidly-progressing disease with a post-diagnosis survival period of 12-15 months and, currently, there are only limited therapies available \cite{cancers6041953}.  Obtaining molecular information of GBM tumors entails an invasive medical procedure on the patient that is costly in terms of both time and resources.  In comparison, magnetic resonance images (MRIs) of these tumors are easily accessible and often readily available.  Being able to effectively utilize MRIs of GBM tumors in computational settings increases the potential for well-developed statistical methodology to have a significant impact in cancer research and future treatment strategies.  

There are two key aims of our work in this paper: first, to quantify GBM tumor images to integrate medical imaging information into statistical models; and second, to explore the utility of medical imaging information in clinical studies of GBM.  To achieve the first aim, we develop a novel statistic, the smooth Euler characteristic transform (SECT), that summarizes shape information of GBM MRIs as a collection of smooth curves.  This allows the direct implementation of existing statistical models from functional data analysis (FDA); in particular, it allows tumor shape information to be used as a covariate in regression frameworks.  To achieve the second aim, we study a cohort of individuals with publicly available MRIs from The Cancer Imaging Archive (TCIA) \cite{Clark:2013,Scarpace:2016aa}, as well as matched genomic and clinical data collected by The Cancer Genome Atlas (TCGA) \cite{TCGA:2008aa}.  Through our extensive predictive analysis, we demonstrate a clinically-relevant connection between the shape of brain malignancies and the variation of survival-based outcomes that are driven by molecular heterogeneity.

The remainder of this paper is organized as follows. In Section \ref{sec:top}, we outline the theoretical concepts used to quantify shape information of tumors and highlight their statistical utility; we also detail the construction of our statistic that summarizes tumor shape information, the SECT. In Section \ref{sec:bfkm}, we detail how regression methodologies for functional covariates are naturally suited to model the curves that capture tumor shape information. This connection with functional data allows us to specify a general regression model that intakes tumor shape information and turns out to be particularly powerful when conducting predictive inference. For our case study, we focus on Gaussian process (GP) regression with Markov chain Monte Carlo (MCMC) inference. In Section \ref{sec:app}, we use the GP modeling framework to predict the clinical outcomes of GBM patients using gene expression data, existing morphometric and volumetric tumor image quantifications, and our proposed tumor shape summaries. Here, we perform a comparative study between each covariate type across different regressions generated by various covariance functions. Finally, in Section \ref{sec:discussion}, we close with a discussion on possible future research.


\section{Quantifying Tumor Images Using Topology} \label{sec:top}

In this section, we develop a summary statistic that captures shape information from MRI images of GBM tumors, which will then be used as covariates in a regression model. The key strategy is to construct these statistics as a function that maps shapes into a Hilbert space. This function has two important properties: (i) it is injective, and (ii) it admits a well-defined inner product structure. Notably, the inner product structure allows us to adapt ideas from functional data analysis to specify general regression models that use shape summary statistics as predictor variables.

\subsection{Background on Summary Statistics for Shape Data} \label{sec:summary}

Classical approaches represent shapes as a collection of landmark points \cite{Kendall:1984aa,Bookstein:1997aa,DrydenMardia98}. This data representation was implemented partly due to the limited image processing technology of the time. Current imaging technologies have since greatly improved and now allow three-dimensional shapes to be represented as meshes, which are collections of vertices, edges, and faces. Figure S1 depicts an example of a mesh representation for a brain tumor and ventricles. Recently, methods have been developed to generate automated geometric morphometrics for mesh representations \cite{Boyer:2011aa,LipmanDaubechies2011,Al_Aifari:2013aa,Boyer:2015aa}. However, despite these advancements, both user-specified and automated landmark-based methods are known to suffer from structural errors when comparing shapes that are highly dissimilar. Some examples of structural errors include: inaccurate pairwise correspondences between landmarks, alignment problems between dissimilar shapes, and global inconsistency of pairwise mappings. These structural errors tend to accumulate as the number of landmarks imposed on each shape increases, and a high number of these points is often required to accurately capture shape information (especially when analyzing diverse shapes) \cite{Gao:2018aa}. Such complications generally make landmark-based approaches less attractive. 

Most recently, an approach known as the persistent homology transform (PHT) was developed to comprehensively address issues induced by landmark-based methods, and to maintain robust quantification performance for highly dissimilar and non-isomorphic shapes \cite{Turner:2014aa}. While the PHT allows for the comparison of shapes without requiring landmarks, it does so by producing a collection of persistence diagrams---multiscale topological summaries used extensively in topological data analysis (TDA). This is restrictive because the geometry of the resulting summary statistics does not allow for an inner product structure that is amenable to (generalized) functional data models \cite{Turner:2014ab}. We propose the smooth Euler characteristic transform (SECT) because it builds upon the theory of the PHT, in that it also produces a topological summary statistic, but it is constructed to be able to integrate shape information in regression-based methods. This proves to be particularly useful in our case study on predicting clinical outcomes in GBM.

\subsection{Homology and Persistence}\label{subsec:top}

We begin by developing an intuition for {\em persistent homology} \cite{Edelsbrunner:2000aa, Zomorodian:2005aa}, which is a foundational concept in TDA. Briefly, persistent homology can be viewed as the data-analytic counterpart to {\em homology}---a theoretical concept from of algebraic topology, where the goal is to study the shape of abstract mathematical objects, such as sets and spaces, by counting occurrences of geometric patterns. In homology, the geometric patterns of interest are holes: homology groups provide a mathematical language for describing and keeping track of holes of an abstract mathematical object. The motivation behind classical algebraic topology is to then use these holes to distinguish between or suggest similarities among different abstract mathematical objects. For a more detailed review and theoretical discussion of these concepts, see the Supplementary Material.

\paragraph{Homology.}

Homology is particularly relevant to our application and case study in GBM. Intuitively, not only does it describe contrasting physical tumor characteristics, but it also implicitly captures some information about the stage of disease progression. For example, {\em necrosis} is a form of cell injury which results in the premature death of cells. {\em Multifocality} is a radiological observation where individual tumor cells separate from the main mass and disperse elsewhere within the brain. From an imaging perspective, necrotic regions show up as dark regions (or holes) within a tumor, while multifocal tumors appear as segregated masses. Examples of both necrosis and multifocality captured by MRI images are shown in Figure \ref{Fig_S2}. It has been suggested that the more necrosis or multifocality there is in a GBM tumor, the more aggressive the disease \cite{pmid8635139,pmid26323991}. Applying homology to radiomic studies not only identifies such phenomena, but also tracks the number of times they occur and thereby provides a notional measure of disease severity.

Homology is indexed by integers: the 0th-degree homology captures the number of connected components in the shape, the 1st-degree homology captures the number of loops, and the 2nd-degree homology captures the number of voids. In the context of our GBM application, degree 0 homology corresponds to tumor masses and lesions. Degrees 1 and 2 homology correspond to necrosis, depending on whether we are analyzing 2-dimensional image slices from an MRI or the 3-dimensional tumor as a whole. 

Despite its intuitive description, computing homology can be challenging. To this end, it is often convenient to represent the shape as a discrete union of simple building blocks, ``glued" together in a combinatorial fashion. An important example of such a building block is the {\em simplex}: simplices are skeletal elements that take the form of vertices, edges, triangles (faces), tetrahedra, and other higher dimensional structures. A {\em simplicial complex} $K$ is a collection of simplices and represents the discretization of a shape or tumor. Meshes that represent three-dimensional shapes are particular examples of finite simplicial complexes (again see Figure \ref{Fig_S1}). There are two key interests in discretizing shapes into simplicial complexes. First, there exist efficient algorithms to compute homology for such discretizations; and second, discretization is essential for applying these abstract concepts to real data, where any given dataset will necessarily be finite. 

In this paper, we use the notation $H_k(K)$ to denote the $k$-th homology group for the simplicial complex $K$. This corresponds to the collection of the $k$-dimensional elements of the simplicial complex. For example, $H_0(K)$ corresponds to the collection of vertices of the simplicial complex or, equivalently, to the collection of connected components of the shape (e.g.~the masses and lesions of a tumor).

\paragraph{Persistent Homology.}

\begin{figure}[H]
\centering
\includegraphics[width = \textwidth]{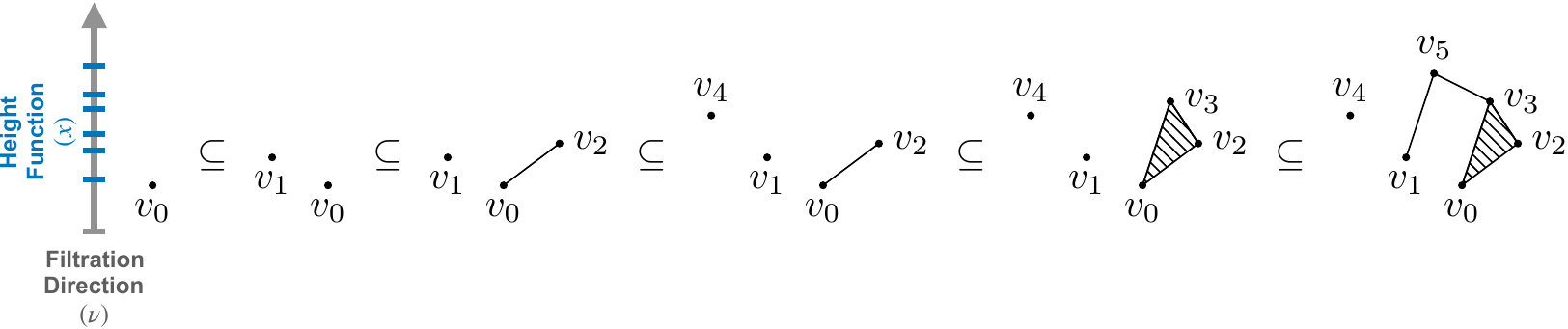}
\caption{Demonstrating a filtration by height (as a function of $x$) in the vertical direction $\nu$ of a simplicial complex $K$. The inclusions ($\subseteq$; from left to right) indicate the evolution of $K$ with the filtration depicted by the hash marks (from the bottom to top) on the $y$-axis, where each element (i.e.~vertex, edge, and face) is included at its maximal height $x$ in the vertical direction $\nu$, as given in Equation (\ref{eqn:height}) from Section \ref{subsec:sect}. Starting from the lefthand side, the vertex $v_0$ is first added and it remains present in each subsequent inclusion, while the vertex $v_1$ emerges during the second inclusion and then remains. Similarly, the edge $v_0v_2$ connecting vertices $v_0$ and $v_2$ appears at the third inclusion, and the face $\langle v_0v_2v_3 \rangle$ forms at the fourth. Altogether, according to these inclusions, the collection of vertices in $K$ are shown to evolve as follows: (i) $\langle v_0 \rangle$; (ii) $\langle v_0, v_1 \rangle$; (iii) $\langle v_0, v_1, v_2 \rangle$; (iv) $\langle v_0, v_1, v_2, v_4 \rangle$; (v) $\langle v_0, v_1, v_2, v_3, v_4 \rangle$; and at the final inclusion, (vi) $\langle v_0, v_1, v_2, v_3, v_4, v_5 \rangle$. Similarly, the collection of edges in $K$ is said to evolve as follows: $\langle v_0v_2 \rangle$ appears at the third, $\langle v_0v_2, v_0v_3, v_2v_3 \rangle$ emerges at the fifth; and $\langle v_0v_2, v_0v_3, v_2v_3, v_1v_5, v_3v_5 \rangle$ forms at the final inclusion. This evolution of vertices, edges, and faces forms (persistence) vector spaces of $i$-chains, and may be written as such for concise notation (see the \ref{appendix:math} for further details). A version of this figure has been previously published \cite{Turner:2014aa}.}
\label{Fig1}
\end{figure}

{\em Persistent homology} applies homology to data by continuously tracking the evolution of homology in the data at different scales (or resolutions). It can thus be seen as a way to extract and summarize geometric information. In persistent homology, the index $s$ of a {\em filtration} tracks the homological evolution. A filtration is a collection of simplicial complexes $\{K_s\}$ where the index $s$ induces totally ordered sets $K_i \subseteq K_j$ for $i < j$. As $s$ increases, the sequence of simplicial complexes $\{K_s\}$ also changes and grows. In this way, the index $s$ of the filtration $\{K_s\}$ tracks the scale according to which the ``shape" of the data changes and grows. The shape information at each scale $s$ is encoded by the homology groups $H_k(K_s)$ of the simplicial complex $K_s$. More specifically, $H_0$ corresponds to the vertices, $H_1$ corresponds to edges, and $H_2$ corresponds to the faces of the simplicial complex or discretized shape. An example of a filtration is depicted in Figure \ref{Fig1}. Here, the index $s$ corresponds to the value of height function (which depends on some variable $x$ and is discussed in detail further below) in the vertical direction $\nu$. We see the evolution of vertices, edges, and a face appearing sequentially with height. Higher order structures are revealed as $s$ increases.

Computing persistent homology produces a collection of intervals for each degree of homology, where each interval represents a $k$-dimensional topological feature (e.g.~a connected component, loop, or void for a general, three-dimensional shape) that is ``born" at the parameter value given by the left endpoint of the interval, and ``dies" at the value at the right endpoint. The length of the interval corresponds to how long the topological feature ``lives," or persists. In this paper, we consider these intervals to be represented by a {\em persistence diagram}. Persistence diagrams treat the start and end points of each interval as an ordered pair, and displays them as plotted points on a plane where the $x$-axis corresponds to birth time and the $y$-axis is the death time. Thus, one can consider a persistence diagram as a collection of points on and above the diagonal, with the set of points on the diagonal having infinite multiplicity (and included for regularity conditions; see the \ref{appendix:math} for further detail).

\paragraph{Persistent Homology Transform.}

The PHT captures shape information by collecting persistence diagrams of all degrees of homology, for all possible orientations of the shape. More formally, for a $d$-dimensional shape, the PHT results in $d$-many persistence diagrams arising from height function filtrations over infinitely-many direction vectors on the surface of the sphere. The space of persistence diagrams is a complicated, but theoretically well-defined probability space \cite{frechet}. In particular, it is a metric space, meaning that distances between persistence diagrams may be defined. This is important because distances between PHT summary statistics provide a way of comparing shapes. The injectivity of the PHT for two- and three-dimensional shapes \cite{Turner:2014aa}, or the one-to-one relation between the shape itself and its infinite collection of persistence diagrams, guarantees that the PHT effectively summarizes all relevant information about the shape.

Considering all possible directions on the surface of the sphere to summarize shape information is particularly well-suited to our radiomics application. MRI scans of the brain are known to be subject to noise: the positioning of patients' heads could vary both between patients and individual scans, causing image registration issues. Considering all directions on the surface of the sphere bypasses this problem, and incorporates perturbations directly into the statistic. This is an important feature of the PHT that we retain in the development of the SECT. We expand upon the PHT to produce a collection of continuous, piecewise linear functions that live in Hilbert space $\mathbb{L}^2$. The corresponding inner product structure inherent to Hilbert spaces allows us to apply the SECT to a much broader set of statistical methodologies. It is worth noting that for select covariance functions, the PHT can be adapted to nonparametric statistical models \cite{Reininghaus:2015aa,Kwitt:2015aa,Kusano:2017aa}, but this class is considerably limited.

\subsection{Smooth Euler Characteristic Transform}\label{subsec:sect}

While the SECT uses the same underlying mathematical principles as the PHT, it produces a collection of continuous, piecewise linear functions rather than persistence diagrams. The SECT implements persistent homology via the Euler characteristic (EC), which is a topological invariant that appears in many branches of mathematics. In terms of homology, the EC counts the ranks of the homology groups (i.e.~the Betti numbers, $\beta_k$, for the $k$-th homology group $H_k$) in an alternating sum and thus reduces the mathematical description of holes in a topological space from an algebraic group structure to an integer.
\begin{definition}
Let $X$ be an arbitrary topological space, $H_k(X)$ be the $k$-th homology group of $X$, and $\beta_k$ be the rank of $H_k(X)$. The Euler characteristic (EC) $\chi(X)$ of $X$ is the alternating sum
\begin{align*}
 \chi(X) = \beta_0 - \beta_1 + \beta_2 - \beta_3 + \cdots =  \sum_{k=0}^{\infty} (-1)^k \beta_k.
\end{align*}
\end{definition}
\noindent For a discretized shape or surface in three dimensions represented as a simplicial complex $K$, the EC may be analogously defined by the number of simplices in $K$ by
\begin{align*}
\chi(K) = V - E + F,
\end{align*}
where $V$, $E$, and $F$ are the numbers of vertices ($0$-simplices), edges ($1$-simplices), and faces ($2$-simplices), respectively.  

Just as homology may be augmented to persistent homology by considering a filtration, ECs may also be calculated with respect to a filtration. The result is an EC curve, which tracks the progression of the EC as a function with respect to the filtration. Let the dimension $d=\{2,3\}$, and fix a direction $\nu$ on the surface of the unit circle or sphere $S^{d-1}$ (where $\nu \in S^{d-1}$). Let $\mathcal{M}_{d-1}$ be the set of all closed, compact subsets (shapes) embedded in $\mathbb{R}^d$ that can be represented in a finite, discrete manner as simplicial complexes \cite{edelsbrunner2010computational}. Next, denote the simplicial complex representation of $M \in \mathcal{M}_{d-1}$ by $K$, and let $K_\nu$ indicate the $\nu$-orientation of $K$.  The {\em sublevel set filtration} of $K_\nu$ parameterized by a height function $r(\bullet, \bullet)$ is the set $\{ x \in K : x \cdot \nu \leq r \}.$ The $\nu$-directional parameter height function $r_\nu(\bullet, \bullet)$ is
\begin{equation}
\begin{aligned}
r : K \times S^{d-1} & \rightarrow \mathbb{R}\label{eqn:height}\\
\{ x, \nu \} & \mapsto x \cdot \nu.
\end{aligned}
\end{equation}
\noindent Denote the extremal heights from this filtration by
\begin{align*}
a_\nu & := \min\{ r_\nu(x), x \in K \},\\
b_\nu & := \max\{ r_\nu(x), x \in K \}.
\end{align*}

\noindent We use the subscript notation to denote the simplicial complex representation $K$ of a shape $M$, in the direction $\nu$, as $K_\nu$ for $d = \{2,3 \}$. Similarly, we use the superscript notation $K_\nu^x$ to denote the varying simplicial complex of $K_\nu$, generated by a sublevel set filtration with respect to Equation \eq{eqn:height} and defined by varying $x \in K_\nu$.

\begin{definition}
\noindent The {\em EC curve} of $K$ (which discretizes $M$) in the direction $\nu$ is defined by
\begin{equation}
\begin{aligned}
\label{eq:ECcurve}
\chi_\nu^K : [a_\nu, b_\nu] & \rightarrow \mathbb{Z} \subset \mathbb{R}\\
x & \mapsto \chi\big( K_\nu^x \big).
\end{aligned}
\end{equation}
\end{definition}

\begin{figure}[H]
\centering
\subfigure[]{
\includegraphics[width=0.31\textwidth]{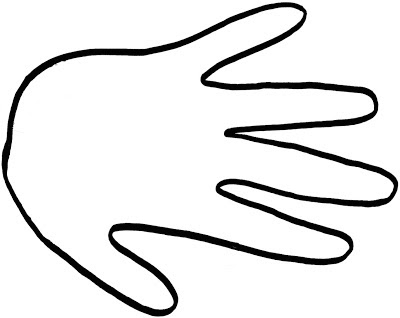}
\label{Fig2A}
}
\subfigure[]{
\includegraphics[width=0.31\textwidth]{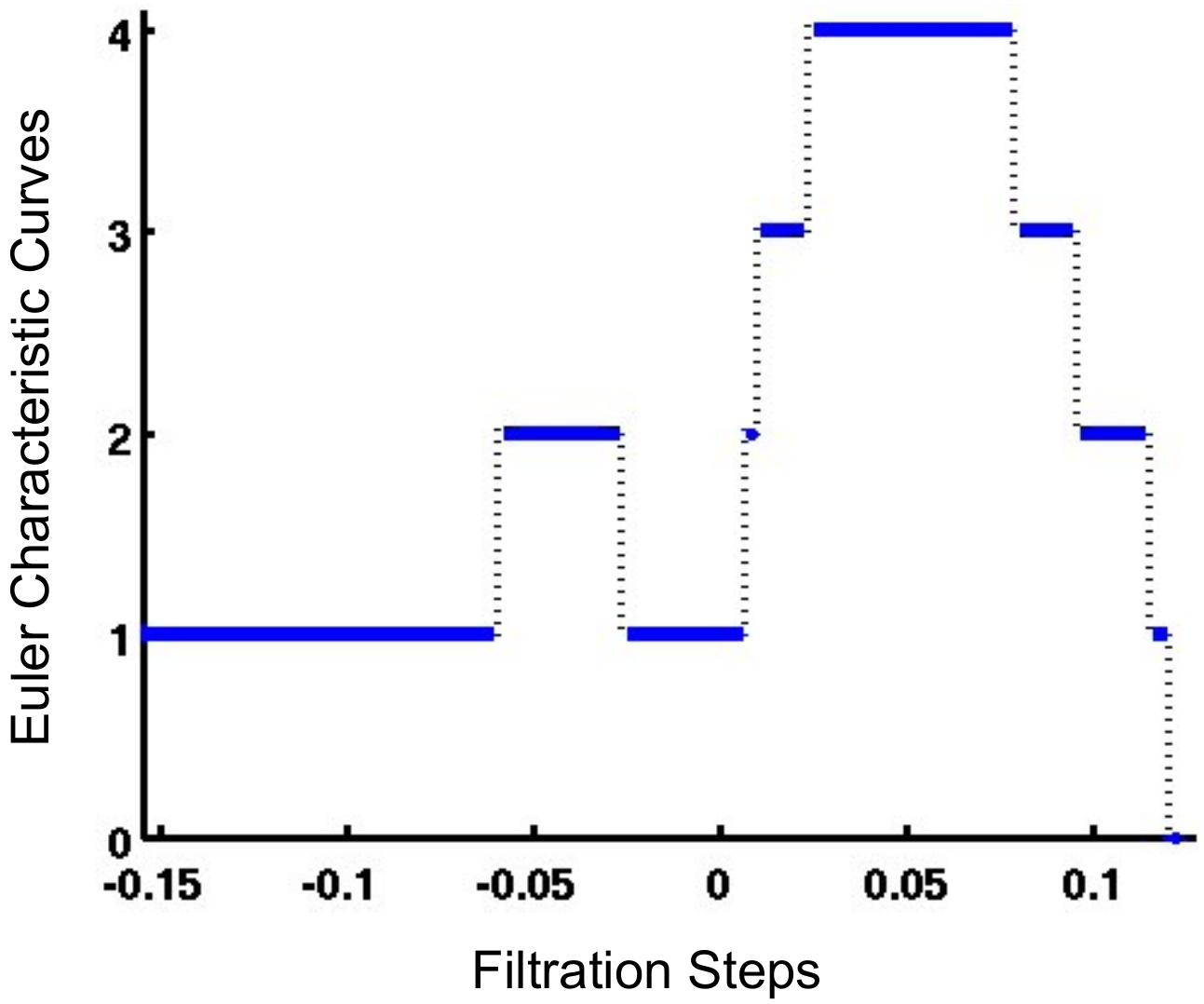}
\label{Fig2B}
}
\subfigure[]{
\includegraphics[width=0.31\textwidth]{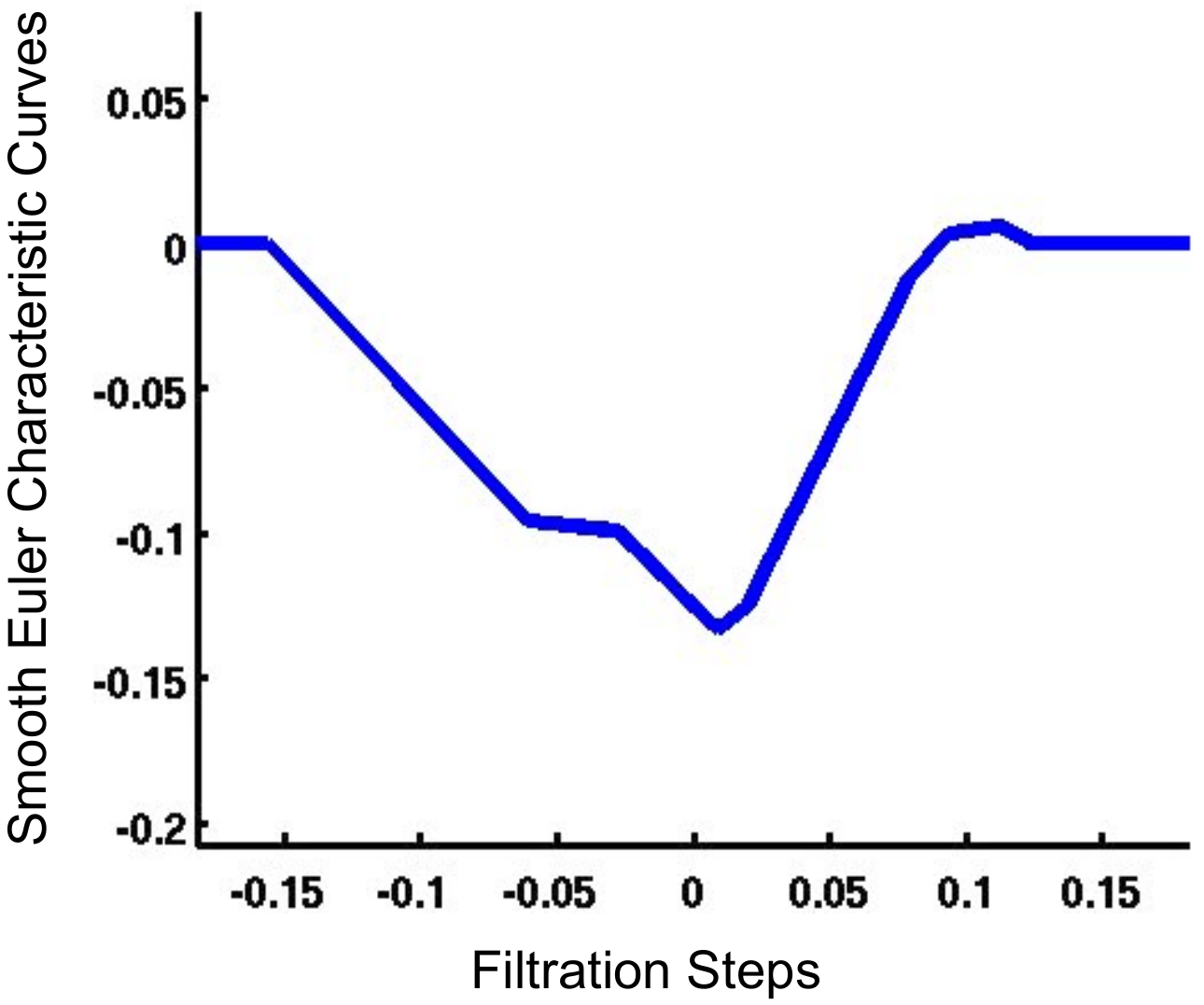}
\label{Fig2C}
}
\caption{Example illustrating the evolutionary tracking of topological features for a given shape. Figure (a) displays a 2D contour of a hand, for which the Euler characteristic (EC) is calculated and tracked with respect to a horizontal filtration. Figure (b) shows the Euler characteristic curve of the 2D contour. At the leftmost filtration level on the $x$-axis, the EC value is equal to 1 on the $y$-axis. This indicates one connected component corresponding to the thumb in the hand. Just before level -$0.05$, the index finger appears, which increases the value of the EC to 2. Figure (c) illustrates the corresponding smooth Euler characteristic (SEC) curve. This found by taking the mean value of the EC curve in (b), subtracting it from every point along the $x$-axis, and integrating over the range of extremal heights (see Section \ref{subsec:sect}). Since there are no holes in a hand, the EC reduces to the sum of connected components as they appear with the filtration. A version of this figure has been previously published \cite{Turner:2014aa}.}
\label{Fig2}
\end{figure}

\noindent The EC curve tracks the evolution of the EC up to (and including) the largest subcomplex of $K_\nu^x$ contained in the sublevel set $r_\nu^{-1}\left( (-\infty, x] \right)$. See Figure \ref{Fig2A} for an illustrative example of the evolution of the EC on the two-dimensional contour of a hand. Here, the direction is the horizontal direction to the right of the $y$-axis. The value of the EC changes as the sweep over the palm first reveals the thumb, and then the separation between the ring and pinky fingers, followed very shortly by the separation between the index and middle fingers, and so on. The EC curve of this filtration is plotted in Figure \ref{Fig2B}. 

The same rotational summary technique of the PHT may be adapted to ECs as follows.

\begin{definition}[Previously in \cite{Turner:2014aa}] In considering a directional sweep over the surface of the sphere $S^{d-1}$, and calculating the corresponding EC curves $\chi^K_\nu$ of the finite simplicial complex representations $K_\nu$ for every direction $\nu \in S^{d-1}$, the {\em Euler characteristic transform (ECT)} is defined as follows:
\begin{equation}
\begin{aligned}
\label{eq:nonsmoothECT}
\text{ECT}(K): S^{d-1} & \rightarrow \mathbb{Z}^{\mathbb{R}}\\
\nu & \mapsto \chi\big( K_\nu \big).
\end{aligned}
\end{equation}
\end{definition}

\noindent In other words, the ECT of a shape collects EC curves of the shape, over all directions on the surface of the sphere.

The EC curve in Equation \eq{eq:ECcurve} and its corresponding ECT in Equation \eq{eq:nonsmoothECT} are piecewise constant, integer-valued functions. These discontinuities can affect the stability of this representation (e.g.~see Figure \ref{Fig2B} where there are sharp jumps in the curve). We therefore propose a reformulation of Equation \eq{eq:nonsmoothECT} that allows for a type of summary that can be used in a wider range of statistical analyses. We do this by smoothing a centered variant of function. The centered variant is given by taking the mean of curve $\bar \chi_\nu^K$ over $[a_\nu, b_\nu]$ and subtracting it from the EC $\chi_\nu^K(x)$ at every $x \in [a_\nu, b_\nu]$. This produces a centered EC curve in the direction $\nu \in S^{d-1}$,
\begin{equation}
\begin{aligned}
Z_\nu^K : [a_\nu, b_\nu] & \rightarrow \mathbb{R}\\
x & \mapsto \chi_\nu^K(x) - \bar \chi_\nu^K.\label{eq:centered_EC}
\end{aligned}
\end{equation}
We set the value of $Z_\nu^K$ to be zero outside the interval $[a_\nu, b_\nu]$ by default. Integrating the curve gives the following smoothed construct.
\begin{definition}
The {\em centered, cumulative Euler characteristic curve} or {\em smooth Euler characteristic curve (SEC)}, for a fixed direction $\nu \in S^{d-1}$, is defined for all $y \in \mathbb{R}$ as
\begin{equation}
\begin{aligned}
\text{SEC}(K)& : \mathbb{R} \rightarrow \mathbb{L}^2\\
F_\nu^K(y) & := \int_{-\infty}^y Z_\nu^K(x)\diff x.\label{eqn:SEC}
\end{aligned}
\end{equation}
\end{definition}

\noindent The SEC is a continuous, piecewise linear function with compact support $[a_\nu, b_\nu]$ by construction. Therefore, it is an element of the Hilbert space $\mathbb{L}^2$ of square integrable functions on $\mathbb{R}$. The counterpart to Figure \ref{Fig2B}, smoothed by the procedure described above resulting in the SEC, is visually illustrated in Figure \ref{Fig2C}. We now formally define the smooth Euler characteristic transform.

\begin{definition}
The {\em smooth Euler characteristic transform (SECT)} for a simplicial complex $K$ of a shape $M \subset \mathbb{R}^d$, with $d=\{2,3\}$, is the map 
\begin{equation}
\begin{aligned}
\text{SECT}(K): S^{d-1} & \rightarrow \mathbb{L}^2[a_\nu, b_\nu]\\
\label{SECT}
\nu & \mapsto F_\nu^K(b_\nu)
\end{aligned}
\end{equation}
for all $\nu \in S^{d-1}$. Each curve $F_\nu^K$ is also an element in the Hilbert space $\mathbb{L}^2$. The following metric can therefore be used to define distances between two simplicial complexes (discrete shape representations) $K_1$ and $K_2$,
\begin{equation}
\label{eqn:dist_SECT}
\mbox{dist}_{\mathcal M_{d-1}}^{\text{SECT}}(K_1,K_2) :=\bigg( \int_{S^{d-1}} \big\| F_\nu^{K_1} - F_\nu^{K_2} \big\|^2 \diff\nu \bigg)^{1/2}.
\end{equation}
\end{definition}

\noindent The advantage of the SECT over the PHT is that SECT summaries are a collection of curves and have a Hilbert space structure. This means that their structure allows for quantitative comparisons using the full scope of functional and nonparametric statistical methodology. The SECT is also an injective map and the following corollary is an immediate consequence of previous results \cite{Turner:2014aa}.

\begin{corollary}
The smooth Euler characteristic transform is injective for two- and three-dimensional shapes, i.e.~when the domain is $\mathcal{M}_{d-1}$ for $d = \{2,3\}$. 
\end{corollary}

\noindent The injective property of the SECT suggests that it concisely summarizes the original shape data. Mathematically, the SECT maps between the space of all shapes with a finite simplicial complex representation $\mathcal{M}_{d-1}$ and the Hilbert space $\mathbb{L}^2$. Thus, injectivity between these two spaces means that for a given SECT statistic in $\mathbb{L}^2$, there is a (unique) corresponding shape with some finite complex representation in $\mathcal{M}_{d-1}$. However, note that enough directions $\nu \in S^{d-1}$ must be taken for this corollary to hold since, for any one fixed direction, it is not true that the EC curve (upon which the SECT construction depends) is injective. An illustration of this fact is depicted in Figure \ref{Fig_S3}. To determine the number of directions to use in practice, we perform a sensitivity analysis with many different combinations of numbers of directions and sublevel sets. In our application of interest and case study, we find prediction results (with SECT features as predictor variables) to be reasonably robust to our final choice of numerical parameters.

\paragraph{Small Note on Information Loss.} 

Based on its homological definition and filtration (defined by a height function), the SECT will always capture all topological and integral geometric (i.e.~size) information about a shape. However, there are instances where information about texture-based features may not be captured in the SECT summary statistic. Intuitively, this is dependent upon the granularity of the filtration defined by the height function. In theory, a continuous height function would mitigate this issue, but this is difficult to implement in practice. As a result, coarse filtrations with too few sublevel sets will cause the SECT to miss or ``step over'' very local undulations in a shape. In the context of our GBM case study, this can occur if a particular tumor is made up of small focal lesions that are collectively important in explaining survival outcomes. For those cases, we would not observe this variance and presumably suffer from predictive performance. 


\section{Functional Regression Models with Tumor Shape Information as Covariates}\label{sec:bfkm}

In the previous section, we formally specified the SECT which allows us to map shapes into a space that: (i) is represented by collection of curves, and (ii) has a well-defined inner product structure. We will now discuss how functional data analysis (FDA) is a particularly suitable framework to specify a general regression model that uses tumor shape information (in the form of topological summary statistics, captured by the SECT) as covariates. The goal of FDA is to model data that are continuous functions (e.g.~curves, response surfaces, or images) \cite{Muller:2005aa,Ferraty:2006aa,Ramsay:2006aa,Muller:2008aa,Pomann:2016aa}. The key idea here is that these functions can be considered as elements in a Hilbert space for which one can specify statistical models using stochastic processes \cite{Preda:2007aa,Morris:2015aa,Kadri:2010aa,Wang:2016aa}. In this paper, we will use a class of stochastic processes that is often referred to as Gaussian processes (GPs) \cite{Wahba,Wolpert:2007aa,Yuan:2010aa}.

\subsection{Gaussian Process Regression}

Denote the shape information (i.e.~SECT representation) of a GBM MRI scan as $\bF(t)=\{F_\nu\}_{\nu=1}^m$ measured over $m$ directions. A functional linear model considers a continuous response variable $\by$ and covariates that are square integrable functions $\bF(t)$ on the real interval $\mathcal{T}$, where $t\in \mathcal{T}$ under the following parametric form \cite{Muller:2005aa},
\begin{align*}
\by = \langle \bF(t),\balpha(t)\rangle+\bvarepsilon, \quad \bvarepsilon\sim\cN(\mathbf{0},\tau^2\bI),
\end{align*}
where the residual noise $\bvarepsilon$ is assumed to follow a multivariate normal distribution with mean zero and scaled variance parameter $\tau^2$, and $\bI$ is used to denote the identity matrix. Notice that similar to traditional linear regression models, $\balpha(t)$ is an unknown smooth parameter function that is now square integrable on the domain $\mathcal{T}$, and $\langle\bullet,\bullet\rangle$ denotes a well-defined inner product. In the context of our radiomic case study, the assumption of a linear relationship between the response variable $\by$ and functional covariates $\bF(t)$ may be too restrictive. For example, when modeling the topological landscape of brain tumors (as we will do in Section \ref{sec:app}), it is reasonable to assume that interactions between modes of brain activity extend well beyond additivity \cite{Friston:2000aa}. As a result, we formulate a general functional regression model that has the flexibility to incorporate possible nonlinear interactions. The methodology we use is Gaussian process (GP) regression.

There are two key characteristics of a GP regression model. The first key element is a positive definite covariance function, $\sigma: \mathbb{L}^2 \times \mathbb{L}^2 \rightarrow \mathbb{R}$, where again $\mathbb{L}^2$ is the Hilbert space of the SECT functional covariates such that $\bF(t) \in \mathbb{L}^2$. The second key element is the reproducing kernel Hilbert space (RKHS) that is induced by the covariance function. Given the eigenfunctions $\{\psi_l\}_{l=1}^\infty$ and eigenvalues $\{\lambda_l\}_{l=1}^\infty$ of the finite integral operator defined by the covariance function \cite{Merc:1909}, we have
\begin{align*}
\int_{\mathcal{T}} \sigma(\bu,\bv)\mathrm{d}(\bu,\bv) <\infty, \quad \quad \lambda_l \psi_l(\bu) = \int_{\mathcal{T}} \sigma(\bu,\bv) \psi_l(\bv) \diff \bv,
\end{align*} 
where $\bu = \bu(t)$ and $\bv = \bv(t)$, and an RKHS can be formally defined as the closure of a linear combination of basis functions $\{\sqrt{\lambda_l}\psi_l(\bv)\}_{l=1}^\infty$ \cite{Wolpert:2007aa}. One may conduct inference in an RKHS by assuming a Gaussian process prior distribution over the functional covariates directly \cite{Rasmussen},
\begin{align}
f\left(\bF_{i}(t)\right)\sim\mathcal{GP}\big(\mu(\bF_{i}(t)),\, \sigma(\bF_{i}(t),\bF_{j}(t)) \big), \quad \quad i,j = 1,\ldots,n\label{GP}
\end{align}
where $f(\bullet)$ is a smooth operator from $\mathbb{L}^2$ to $\mathbb{R}$ that is completely specified by its mean function and positive definite covariance function, $\mu(\bullet)$ and $\sigma(\bullet,\bullet)$, respectively. Recall from Section \ref{sec:top} that, in practical applications, there are a finite number of observed topological summary statistics taken from a given geometric object. Therefore, if we condition on these finite set of locations, the prior distribution in \eq{GP} may be represented as multivariate normal \cite{Kolmogorov:1960aa}. Consider the following joint ``weight-space'' probabilistic regression model to complete our specification \cite{Rasmussen},
\begin{align}
\by = \bm{f} +\bvarepsilon, \quad \bm{f}\sim\cN(\mathbf{0},\Sigma(\bF(t),\bF(t))), \quad \bvarepsilon\sim\cN(\mathbf{0},\tau^2\bI),\label{mod4}
\end{align}
where $\bm{f}=\big[f(\bF_{1}(t)),\ldots,f(\bF_{n}(t)) \big]^{\T}$ is now assumed to come from a multivariate normal with mean $\mathbf{0}$ (for simplicity) and covariance matrix $\Sigma(\bF(t),\bF(t))$.

\subsection{Posterior Predictive Inference}

We now formally describe how to conduct posterior predictive inferences on clinical phenotypic traits for unobserved patients. Assume that we have received a set of new brain tumors and have computed their corresponding topological summary statistics $\bF^*(t)$. Under the prior in Equation \eq{GP}, we can write the joint distribution between the observed patient responses ($\by$) and the function values taken at the test images ($\bm{f}^*$) as
\begin{align}
\beginmat{c}\by \\ \bm{f}^* \endmat \sim \cN\left(\bm{0}, \begin{bmatrix} \Sigma(\bF(t),\bF(t)) + \tau^2\bI & \Sigma(\bF(t),\bF^*(t)) \\ \Sigma(\bF^*(t),\bF(t)) & \Sigma(\bF^*(t),\bF^*(t)) \end{bmatrix}\right).\label{joint}
\end{align}
Intuitively, if we train the model on $n$ tumors and there are $n^*$ test images, then $\Sigma(\bF(t),\bF^*(t))$ results in an $n\times n^*$ matrix of covariances between each of the training and testing points. Similar interpretations also hold for other covariance entries as well. Deriving the conditional distributions for Equation \eq{joint} then results in a multivariate normal posterior predictive distribution for the test shape smooth operators $\bm{f}^*\cond \by \sim \cN(\bmu^*,\bSigma^*)$, where
\begin{equation}
\begin{aligned}
\bmu^* &= \Sigma(\bF^*(t),\bF(t))\left[\Sigma(\bF(t),\bF(t))+\tau^2\bI\right]^{-1}\by\\ 
\bSigma^* &= \Sigma(\bF^*(t),\bF^*(t)) - \Sigma(\bF^*(t),\bF(t))\left[\Sigma(\bF(t),\bF(t))+\tau^2\bI\right]^{-1}\Sigma(\bF(t),\bF^*(t)).
\end{aligned}
\end{equation}
Note that in many applications, covariance functions can be indexed by a bandwidth or length-scale parameter $\theta$, $\sigma_\theta(\bu,\bv)$. For example, the Gaussian kernel can be specified as $\sigma_\theta(\bu,\bv) = \exp\{-\|\bu-\bv\|^2/2\theta\}$. This bandwidth parameter can be inferred; however, posterior inference over $\theta$ is slow, complicated, and often mixes poorly \cite{Liang:2009aa}. For simplicity, we will work with a fixed bandwidth that is chosen via 10-fold cross validation. 


\section{Predicting Clinical Outcomes in Glioblastoma}\label{sec:app}

To fully illustrate the statistical utility of tumor images (captured by the SECT topological summary statistic), we apply the GP regression model to a GBM radiomic study with two measured clinical outcomes: disease free survival (DFS) and overall survival (OS). 
Some recent work in radiomics has confirmed the utility of imaging data in GBM research. These efforts suggest that the inclusion of shape information improves both the prediction of patient survival outcomes, as well as the classification of tumor subtypes \cite{Gutman:2013aa,Mazurowski:2013aa,Gevaert2014,Macyszyn:2016aa}. It is important to distinguish, however, that most of these previous studies were limited to gross spatial features of cancer images (e.g.~the presence of multifocal tumors, the location of recurrent lesions, or crude volumetric calculations). The SECT offers a novel contribution to radiomic research as a topological representation of imaging data. In this section, we will specifically assess whether topological features are better predictors of DFS and OS prognoses than three other tumor characteristics: (i) gene expression, (ii) tumor morphometry, and (iii) tumor geometry.

\subsection{Genomic and Radiomic Data}\label{subsec:images}

Magnetic resonance images (MRIs) of primary GBM tumors were collected from $n = 48$ patients archived by The Cancer Imaging Archive (TCIA) \cite{Clark:2013,Scarpace:2016aa}, which is a publicly accessible data repository containing medical images of cancer patients with matched genomic and clinical data collected by The Cancer Genome Atlas (TCGA) \cite{TCGA:2008aa}. These patients were selected based on two sets of criteria, namely: (i) individuals had post-contrast T1 axial (transverse) MRIs taken at the time of their diagnosis, and (ii) these patients have matching (mRNA) gene expression data and clinical correlates (e.g.~recorded DFS and OS) that are publicly available \cite{Gao:2013aa,Grossman:2016aa}. There are three key factors that influenced our decision to use this particular subset of samples. First, the T1-weighted MRI with gadolinium contrast is one of the most commonly-used imaging modalities and is often implemented to assay lesions with vascular activity \cite{Adin:2015aa}. Second, axial (transverse) slices were considered as this was the most common representation in TCIA database and resulted in the dataset with the most observations. Third, exclusively using MRIs taken at the time of diagnosis allowed us to avoid any potential confounding factors related to treatment-specific effects that may alter postoperative imaging and/or genomic profiles \cite{Macyszyn:2016aa}.

Each collection of patient MRIs consisted of approximately 23-25 segmented slices of two-dimensional grayscale images (with the exact numbers varying between patients). We segment these images with the computer-assisted program MITKats to extract tumor lesions from the surrounding brain tissue \cite{mitkats}. Briefly, this algorithm first converts MRI images to a grayscale, and then thresholds to generate binary images. Morphological segmentation is then applied to delineate connected components. This is done by selecting contours corresponding to enhanced tumor lesions, which are lighter than healthy brain tissue. For instance, necrosis (or $H_1$ or $H_2$ homology as described above in Section \ref{subsec:top}) presents as dark regions nested within the indicated lesion. An example of a raw image obtained from TCIA, along with its corresponding segmentation, is given in Figure \ref{Fig_S4}.

\begin{figure}[H]
\centering
\subfigure{
\includegraphics[width = 0.41\textwidth]{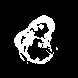}
\label{Fig3A}
}
\subfigure{
\includegraphics[width = 0.49\textwidth]{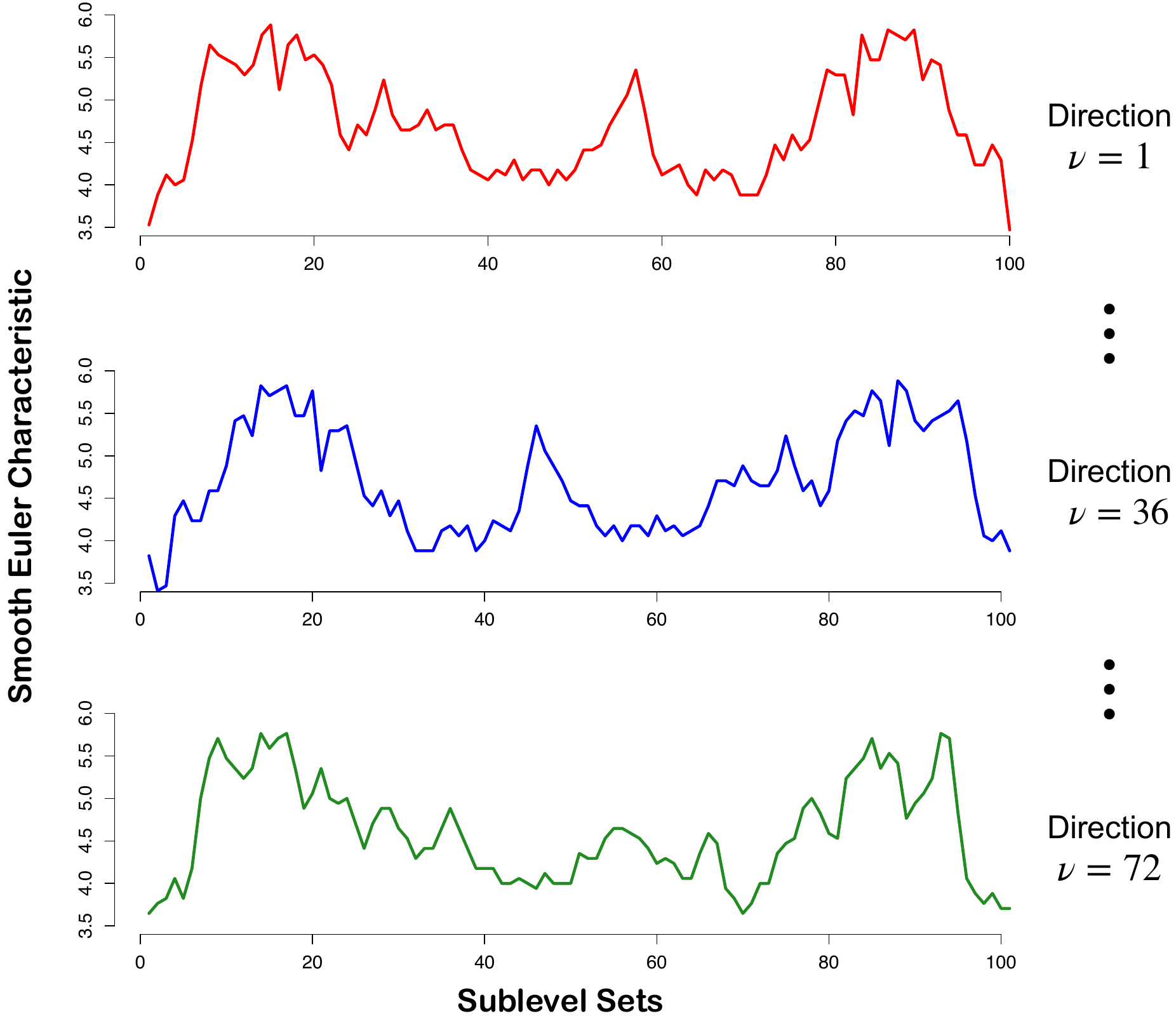}
\label{Fig3B}
}
\caption{Example of a segmented tumor (on the left) with its corresponding SECT curves (on the right). On the right side, we illustrate the curves for directions $\nu =$ 1, 36, and 72, respectively. The $x$-axis shows the number of sublevel sets (i.e.~steps) as the filtration progresses, which we fix to be 100 per slice. In the radiomic prediction analysis, all 72 curves are concatenated together to create a 7200-dimensional covariate vector for the patient.}
\label{Fig3}
\end{figure}

From these segmented images, we collect three types of tumor shape information: morphometric features, geometric measurements, and topological summary statistics. Here, we use the same morphometric features outlined in previous imaging studies \cite{Han:2010aa,Chang:2011aa} (see listed references for specific details on extraction and computation). The final data set consisted of these 212 morphometric predictors corresponding to shape and texture, including cellularity skewness, cytoplasm intensity, nucleus texture, nucleus curvature, and median edge length. We also consider 5 tumor geometric measures. The first is the enhancing volume for each MRI slice, which is summed over lesions in the multifocal case. The other geometric measurements are the core volume of the enhancing and necrotic regions, the longest lesion diameter, and the shape factor of the tumor. For the purposes of this study, we define the shape factor to be the longest lesion diameter divided by the diameter of a sphere with the same volume. Lastly, when computing topological summary statistics, we use 100 sublevel sets per slice and compute a different EC curve for 72 directions evenly sampled over the interval $[0,2\pi]$. After concatenation, this results in 7200-dimensional vectors for each patient. Averaging over these curves for each direction gives the proposed smooth EC statistics. It is well known that reconstructing 3D brain tissue (and corresponding tumors) from 2D slices is a nontrivial task \cite{Cline:1987aa,Amruta:2010aa,Jaffar:2012aa}. Moreover, in the context of our case study, it is not guaranteed that the space in-between individual slices will be the same for each patient. Therefore, we aggregate topological summaries across slices for each direction as a proxy for rotating (accurate) 3D representations of each tumor. Example SECT summary statistics for a segmented tumor are depicted in Figure \ref{Fig3}.

Lastly, we use matched mRNA gene expression levels of the preselected TCGA samples as a baseline data source.  Following specific preprocessing steps from other genomic studies \cite{Singleton:2017aa}, we use the robust multiarray average (RMA) normalization procedure to correct for potential lab-based batch effects and other potential confounders \cite{Irizarry:2003aa}. This resulted in a final data set consisting of 8725 genes, which also passed a pre-specified hybridization accuracy threshold and showed reasonably varying expression across the assay.

\subsection{Prediction Results}

We now compare the each data type's ability to predict two clinical outcomes: disease free survival (DFS) and overall survival (OS). Briefly, DFS is the period after a successful treatment during which there are no signs or symptoms of the cancer, while OS is tabulated as the entire period after the initial treatment where the patient is still alive. It is worth noting that DFS is more commonly used over OS in adjuvant cancer clinical trials because it offers earlier presentation of data \cite{Sargent:2005aa}. This stems from the idea that events due to disease recurrence occur earlier than death from disease, thus resulting in a cleaner signal \cite{Birgisson:2011aa}.

Each tumor feature type is modeled with the GP regression model detailed in Section \ref{sec:bfkm}. In the context of this application, every patient in the data has an official death time. Hence, there is no need for a right-censored analysis or Cox-based methods \cite{Rutledge:2013aa,Fatai:2018aa,Kundu:2018aa}. We use two types of metrics to assess prediction accuracy: (i) the squared correlation coefficient ($R^2$), and (ii) the frequency for which a given data type exhibits the greatest $R^2$, which we denote as Optimal\%. When analyzing each outcome, we randomly split the data 1000 different times into 80\% training and 20\% out-of-sample test sets. In each case, the survival times are centered and scaled to have mean 0 and variance 1 to facilitate the interpretation of results. In order to illustrate the robustness of the SECT, we apply the GP regression framework using three different covariance functions. The goal is to show that the power of the SECT summary statistic is robust to this choice. Here, suppose that $\bu$ and $\bv$ are two different covariate vectors. We consider the linear (gram) kernel $\sigma(\bu,\bv) = \bu^{\T}\bv/p$, where $p$ is the length of the feature vector for a given data type; the Gaussian kernel $\sigma_\theta(\bu,\bv) = \exp\{-\|\bu-\bv\|^2/2\theta\}$; and the Cauchy kernel $\sigma_\theta(\bu,\bv) = (1+\theta\|\bu-\bv\|^2)^{-1}$. As previously mentioned, the last two functions are indexed by a bandwidth or length-scale parameter $\theta$, which we select via 10-fold cross-validation over the grid [0.1,10] with step sizes equal to 0.1. Briefly, a value of 0 denotes a rigid function, while 10 represents a smoothed estimator. In Table \ref{Tab1}, we present the mean $R^2$ and corresponding standard errors across testing splits to show how each tumor characteristic performs while taking into account variability. This table also lists the estimated bandwidths that generated these results.

Overall, our study shows that SECT topological summaries result in the most accurate predictions for survival---particularly for DFS. For example, using the Cauchy kernel function with SECT features resulted in the greatest $R^2$ for both DFS and OS at 0.237 and 0.158, respectively. This led to the SECT being the optimal tumor characteristic 40.5\% of the time for DFS and 36.5\% of the time for OS. There are a few possible explanations for these results. First and foremost, gene expression is known to be highly variable, particularly in GBM \cite{Verhaak:2010aa}. Second, volumetric-based measurements only detail information about tumor size, but this information is likely not enough to be an effective predictor of patient survival on its own. Instead, one could imagine the geometry of a tumor being more useful when paired with the spatial location of lesions; hence, better detailing the severity of the shape. Conversely, morphometric features describe more focal-based characteristics of malignancies. This attention to texture causes more global information about the tumor to be missed. 

These predictive results may also be due to the nature of the clinical outcomes that we chose to model. As previously mentioned, DFS is a prognostic measure of cancer recurrence and corresponds to the reappearance of the disease after initial treatment. This correlate can often be better defined than OS, where the cause of a patient's death may not necessarily be due to cancer-based complications. Indeed, each of the tumor characteristics that we consider generally perform better when predicting DFS versus OS (see Table \ref{Tab1}). Nonetheless, measurements that provided detailed information about one particular aspect of the disease (e.g.~volumetric and morphometric quantities) are not as relevant as those that aim to illustrate a more comprehensive view. Thus, topological features are effective predictors as they provide some notion about both the size and texture of tumors.

\subsection{MRI Specific Consequences on Results}

In our application, the robustness of the SECT to choice of metric is particularly relevant because the geometric structure of the brain is known to be fibrous---meaning that the brain is made up of, and connected by, cerebral fiber pathways \cite{Wedeen:2012aa}. This brings into question the validity of assuming the usual Euclidean metric when quantifying shape. Both volumetric and morphometric analyses require the specification of a metric and, in the case where the usual assumption of a Euclidean measure does not apply, an appropriate one must be constructed. This is not always a straightforward task. Moreover, in fibrous settings, there is also the possibility for the further requirement of defining a geodesic. Examining topological properties, as opposed to metric-based properties, bypasses these technical difficulties. Altogether, incorporating a topological measure that is not based on a metric results in the flexibility to compare tumors of different sizes more seamlessly. Subsequently, this also implicitly allows for comparisons between different stages of the disease without needing to account for time of progression. We hypothesize that these flexible characteristics also contribute to the SECT being a better predictor of prognosis and survival.

\subsection{More Biological Implications}

One key implication from our results in DFS is that there possibly exist correlations between the topology of tumors and the molecular heterogeneity arising from the activation of different recurrence mechanisms. An example of this relationship occurs in (multifocal) tumors where lesions on opposite hemispheres of the brain originate from the same oncogenic effects, but events like therapeutic resistance or cancer recurrence happen in only one hemisphere. This variation can be clinically relevant. It was recently proposed that these type of topologically-based traits are linked to the mutation status of certain oncogenic relapse drivers \cite{Lee:2017aa}. Hence, there is growing evidence that potential pathways of progression in GBM should go beyond the simple consideration of physical proximity (i.e.~closeness in a geometric sense). For instance, a particular path to recurrence in GBM may be due to ambient effects inherent to a particular hemisphere of the brain. The prediction results we present in this work suggest that the topological features extracted by the SECT may be better than simple geometric summaries at providing insight into biological phenomena at the molecular level.


\section{Discussion}\label{sec:discussion}

In this paper, we sought to quantify images of GBM tumors given by MRIs for statistical analyses and to demonstrate the clinical relevance of this information.  To this end, we developed a topological summary statistic transform which maps shapes into a space that admits an inner product structure that is amenable to standard functional and nonlinear regression models. We then used our summary transform to predict the survival of GBM patients using our specified functional Gaussian process regression model. In this study, we compared the predictive accuracy using both molecular biomarkers and shape covariates. The SECT was shown to explain more of the variance in DFS of patients than all other covariates in a wide variety of models defined by various kernel functions. For the Gaussian and Cauchy kernels, in particular, the SECT outperformed the other measures in accounting for the variance in both DFS and OS.

Despite these results, several interesting future directions and open questions still remain. For example, in the current study, we focus solely on measuring how well topological features predict survival. Many studies in the radiomics space use deep learning approaches for accurate classification and prediction-based tasks \cite{Lao:2017aa,Li:2017aa,Bibault:2018aa,Rathore:2018aa}. Unfortunately, in this work, we did not have access to data with large enough sample sizes for the effective training of neural networks. However, in the future, it would be useful to see how our topological summary statistics may be integrated within deep learning frameworks. To ensure power, utilizing protected data from current consortium studies with a large number of participants would be of high interest \cite{Mueller:2005aa,Gounder:2015aa,Sudlow:2015aa}. Moving away from prediction, it would also be useful to infer which particular spatial regions of the tumor are most relevant to clinical outcomes. Recent variable selection approaches for kernel-based methods can be used to infer the directions and segments of the Euler curves that are most relevant \cite{crawford2018bayesian,crawford2018variable}. In this case, an important open problem is having the ability to recover, or partially reconstruct, a shape based on significant SECT summary statistics. Similarly, the distance measure for the SECT stated in Equation \eq{eqn:dist_SECT} provides a framework for comparing the shapes of tumors, and correlating geometric properties with molecular and clinical features. Understanding the relationship between therapeutic strategies, signaling pathway dependence, and tumor shapes would provide useful information about different forms of GBM and their etiologies. We conjecture that greater general knowledge about tumor shape may help in distinguishing true progression from pseudoprogression. Here, progression refers to the growth of the tumor itself, while a pseudoprogressing tumor has been infiltrated by immune cells and other factors.


\section*{Data Availability}

The results shown here are in whole or part based upon data generated by the TCGA Research Network (\url{http://cancergenome.nih.gov/}). DICOM formatted MRI scans and patient clinical information were taken directly from the TCIA web portal (\url{https://wiki.cancerimagingarchive.net/display/Public/TCGA-GBM}). Matched molecular data were downloaded directly from the Genomic Data Commons (GDC) by selecting the RNA-Seq tab option (\url{https://portal.gdc.cancer.gov/projects/TCGA-GBM}). Shape-based summary statistics necessary for replicating this study (i.e.~the segmented tumor images, the volumetric measurements, morphometric data, and topological summary statistics) are also publicly available on the SECT GitHub repository.

\section*{Software Availability}

Software to compute the SECT from images and fit the Gaussian process regression model is publicly available in both R and MATLAB code, and located on the repository \url{https://github.com/lorinanthony/SECT}. The MRI images were segmented using the Medical Imaging Interaction Toolkit with augmented tools for segmentation (MITKats), which was written C++ and is located at \url{https://github.com/RabadanLab/MITKats} \cite{mitkats}.


\section*{Acknowledgements}

During some of this work, LC, AM, and RR~were supported by the National Cancer Institute Physical Sciences--Oncology Network (NCI PS--ON) under Grant No.~5 U54 CA 193313-02. AM was the PI on Pilot Grant Subaward No.~G11124 for research on radiomics and radiogenomics.  AM is also supported by the Irving Institute's CaMPR inititative under Grant No.~GG011557, and would like to acknowledge the support of the New Frontiers in Research Fund--Fonds Nouvelles fronti\`{e}res en recherche (SSHRC--NFRF--FNFR Government of Canada) NFRFE-2018-00431. LC would like to acknowledge the support of grants P20GM109035 (COBRE Center for Computational Biology of Human Disease; PI Rand) and P20GM103645 (COBRE Center for Central Nervous; PI Sanes) from the NIH NIGMS, 2U10CA180794-06 from the NIH NCI and the Dana Farber Cancer Institute (PIs Gray and Gatsonis), and an Alfred P. Sloan Research Fellowship (No.~FG-2019-11622). AXC would like to acknowledge support by the Columbia University Medical Scientist Training Program (MSTP). SM would like to acknowledge funding from NSF DEB-1840223, NIH R01 DK116187-01, HFSP RGP0051/2017, NSF DMS 17-13012, and NSF DMS 16-13261. This work used a high-performance computing facility partially supported by grant 2016-IDG-1013 (``HARDAC+: Reproducible HPC for Next-generation Genomics'') from the North Carolina Biotechnology Center. The authors wish to thank Mao Li (Donald Danforth Plant Science Center) and Christoph Hellmayr (Duke University) for help with the formulation of code, as well as Francesco Abate (McKinsey \& Co.), Katharine Turner (Australian National University), and Jiguang Wang (Hong Kong University of Science and Technology) for helpful conversations and input on a previous version of the manuscript. The authors would also like to acknowledge The Cancer Imaging Archive (TCIA) and The Cancer Genome Atlas (TCGA) initiatives for making the imaging and the clinical data used in this study publicly available. Any opinions, findings, and conclusions or recommendations expressed in this material are those of the author(s) and do not necessarily reflect the views of any of the funders.


\section*{Author Contributions Statement}

LC, AM, SM, and RR conceived the study. LC and AM developed the methods. LC, AM, and AXC developed the algorithms. LC implemented the software and performed the analyses. All authors wrote and revised the manuscript.


\section*{Competing Financial Interests}

The authors have declared that no competing interests exist.


\clearpage
\newpage
\appendix
\renewcommand{\thesection}{Appendix}
\renewcommand{\thesubsection}{A\arabic{subsection}}
\renewcommand{\theAppDefinition}{A\arabic{AppDefinition}}

\setcounter{equation}{0}

\makeatletter 
\renewcommand{\theequation}{A\@arabic\c@equation}

\section{}\label{appendix:math}

In this section, we provide more formal details on the mathematics underlying the concepts of persistent homology and topological data analysis (TDA) for shapes and images. For a complete discussion on theory in TDA and applied topology, the interested reader may refer to a detailed literature (e.g.~\cite{Ghrist2008,Carlsson2009,carlsson2014}).

\subsection*{Homology, Simplicial Complexes, and Persistence}\label{subsec:homology}\addcontentsline{toc}{subsection}{Homology, Simplicial Complexes, and Persistence}

Homology groups provide an algebraic structure to study holes in a topological space. Such holes are captured indirectly by considering what surrounds them. In other words, homology is concerned with studying the boundaries of holes. The fundamental property underlying homology is that the boundary of a boundary is necessarily zero. Algebraicity of such a study refers to group operations and maps that relate topologically-meaningful subsets of a space with one another. In discretizing a general topological space in terms of simplices, and thus studying its simplicial homology, the underlying object of study is a simplicial complex. This is the context from which we work in this paper, and what will be described in detail in this section.

A \emph{$k$-simplex} is the convex hull of $k+1$ affine independent points $v_0,v_1, \ldots v_k$, and is denoted by $\sigma = [v_0, v_1,\ldots, v_k]$. Examples of $k$-simplices are points, lines, and triangles. The $0$-simplex $[v_0]$ is the vertex $v_0$; the $1$-simplex $[v_0,v_1]$ is the edge between the vertices $v_0$ and $v_1$; and the $2$-simplex $[v_0, v_1, v_2]$ is the triangle bordered by the edges $[v_0,v_1]$, $[v_1, v_2]$ and $[v_0, v_2]$. 

\begin{AppDefinition}
\label{def:simp_comp}
A geometric simplicial complex $K$ is a countable set of simplices such that:
\begin{enumerate}[1.]
\item Every face of a simplex in $K$ is also in $K$;
\item If two $k$-simplices $\sigma_1,\sigma_2$ are in $K$, then their intersection is either empty or a face of both $\sigma_1$ and $\sigma_2$.
\end{enumerate}
\end{AppDefinition}

\noindent Fix a dimension $k$ and a field $\mathbb{F}$. Given a shape $M$ with a finite simplicial complex representation (mesh) $K$, a \emph{simplicial $k$-chain} is a linear combination of $k$-simplices $\sum_k c_k \sigma_k$, where $c_k \in \mathbb{F}$ and $\sigma_k \in K$. Here, the sum is taken over all possible $k$-simplices. Denote the set of all such $k$-chains by $C_k(K)$. These $k$-chains may be added (given $c = \sum_k c_k \sigma_k$ and $d = \sum_k d_k \sigma_k$, with $c + d := \sum_k (c_k + d_k) \sigma_k$) and multiplied by scalars. Thus, $C_k(K)$ is a vector space over $\mathbb{F}$ of $k$-chains in $K$, and the set of $k$-simplices forms a canonical basis for $C_k(K)$.

The theory and results developed in this paper rely on the simplifying assumption that $\mathbb{F}$ is the binary field $\mathbb{Z}_2 = \mathbb{Z}/2\mathbb{Z}$. In this case, a $k$-chain is a collection of $k$-simplices, and the {\em boundary} of a $k$-simplex is the sum of its $(k-1)$-dimensional faces. Let $\sigma = [v_0, v_1, \ldots, v_k]$ denote the simplex spanned by the specified vertices. The {\em boundary map} $\partial_k:C_k(K) \to C_{k-1}(K)$ maps a $k$-chain to a $(k-1)$-chain, and is given by 
\begin{align*}
\partial_k\big([v_0, v_1, \ldots, v_k]\big) = \sum_{j=0}^k [v_0,\ldots, v_{-j}, \ldots, v_k]
\end{align*}
with linear extension, where $v_{-j}$ specifies that the $j$-th element is dropped. Elements of $B_k(K) := \operatorname{im} \partial_{k+1}$ are called {\em boundaries}, and elements of $Z_k(K) := \operatorname{ker} \partial_k$ are called {\em cycles}. 

\begin{AppDefinition}
The $k$-th homology group of $M$ is defined by the quotient group 
\begin{align*}
H_k(K):=Z_k(K)/B_k(K).
\end{align*}
\end{AppDefinition}

\noindent The intuition behind a homology group is that it contains information about the structure of $K$. The zeroth homology group $H_0(X)$ is generated by elements that represent connected components of $X$. For example, if $X$ has three connected components, then $H_0(X) \cong \mathbb{Z}_2 \oplus \mathbb{Z}_2 \oplus \mathbb{Z}_2$, where $\cong$ here denotes group isomorphism. For $k\ge 1$, the $k$-th homology group $H_k(X)$ is generated by elements representing $k$-dimensional ``holes" or ``loops" in $X$. A $k$-dimensional hole can be thought of as the result of taking the boundary of a $(k+1)$-dimensional body. The ranks of the homology groups (i.e.~the number of generators) are called the {\em Betti numbers}, and are denoted by $\beta_k(X)  := \mbox{rank}\big(H_k(X) \big)$. The notation $H_*(X)$ refers to all the homology groups simultaneously. In Figure \ref{Fig_S5}, we display the homology of a torus constructed from a simplicial complex.

\subsection*{Persistent Homology}\addcontentsline{toc}{subsection}{Persistent Homology}

A {\em filtration} of simpicial complexes $\mathcal{K}$ is an indexed, nested family of spaces $\mathcal{K} = \{K_s\}_{s=a}^b$ such that $K_{s_1} \subseteq K_{s_2}$ if $s_1 < s_2$. As the parameter $s$ increases, the homology of the spaces $K_s$ may change (e.g.~components are added and merged, cycles are formed and filled up). Examples of two different filtrations are given in Figures \ref{Fig1} and \ref{Fig_S6}. The former illustrates a filtration by height function where the topological structure, in terms of simplicial homology, is revealed at the level of maximal height as $s = x$ increases in the vertical direction $\nu$. The latter illustrates a filtration by radius: the sample points from the space are taken to be centers of balls while the radius $s = \epsilon$ grows from $0$ to $\infty$. Overlapping balls are replaced by a $k$-simplex, depending on the degree of overlap: two overlapping balls are replaced by an edge, three overlapping balls are replaced by a face or triangle, and so on. Formally known as the Vietoris--Rips filtration, this procedure reveals the topological structure in terms of simplicial homology.

The \emph{persistent homology} of $\mathcal K$ is denoted by $\mbox{PH}_*(\mathcal{K})$ and keeps track of the progression of homology groups generated by the filtration. More specifically, the persistent homology contains the information about the homology of the individual spaces $\{K_s\}$, as well as the mappings between the homology of $K_{s_{1}}$ and $K_{s_{2}}$ for every $s_{1} < s_{2}$. Note that persistent homology is also equivalently referred to as {\em persistence}.

\begin{AppDefinition}
Let $K$ be a filtered simplicial complex with $K_1 \subset K_2 \subset \cdots \subset K_S = K$. The $k$-th persistence module derived in homology, or $k$-th persistent homology, of $K$ is
\begin{align*}
\mathrm{PH}_k(K) := \big\{ H_k(K_s) \big\}_{1 \leq s \leq S} \mbox{~with~} \{ \varphi_{s_{1},s_{2}} \}_{1 \leq s_1 \leq s_2 \leq S},
\end{align*}
where each linear map $\varphi_{s_{1},s_{2}}: H_k(K_{s_1}) \rightarrow H_k(K_{s_2})$ is induced by the inclusion $K_{s_1} \hookrightarrow K_{s_2}$ for all $s_1, s_2 \in [1, S]$ with $s_1 \leq s_2$.
\end{AppDefinition}

\noindent Intuitively, the main idea behind persistent homology is to study homology across multiple scales. Rather than restricting ourselves to only one instance of a space, in persistent homology, we study the evolution of the topological structure over a filtration of the entire space. This amounts to beginning with a rigid proximity rule connecting observed data points, and then continuously relaxing this rule whilst studying the corresponding topological progression. 

\subsection*{Barcodes and Persistence Diagrams}\addcontentsline{toc}{subsection}{Barcodes and Persistence Diagrams}

The persistence of the data is encoded in parameterizations of homology groups known as {\em barcodes}---collections of intervals corresponding to the lifetimes of topological features. The left endpoint of a bar is the \emph{birth time} of an element in $\mbox{PH}_*({\mathcal K})$ and can be thought of as the value of $s$ where this element appears for the first time. Conversely, the right endpoint of a bar is the \emph{death time} and represents the value of $s$ where an element vanishes, or merges with another existing element. The convention in studying merging features is to retain and treat the feature that appeared first as if it were continuing its existence beyond the merge event.

Figure \ref{Fig_S6} illustrates the idea behind persistent homology and provides an example of a barcode \cite{Ghrist2008}. In this example, $\epsilon$ corresponds to the filtration parameter value. The filtration is illustrated in the upper panels in the evolution of the simplicial complex in terms of vertices, edges, and faces that are formed with the progression of $\epsilon$. Here, the $H_0$ homology captures connected components; $H_1$ captures cycles whose boundaries are formed by edges between vertices; and $H_2$ captures cycles with boundaries formed by faces. The dashed lines extending from the panels capture instances of the filtration, and link to the bars representing the topological features at particular values of $\epsilon$. As $\epsilon$ progresses, we see connected components merge, cycles form, and fill up. The barcode summarizes the lifetimes of all topological features, classified by their homology groups, in this process. 

Barcodes can thus be considered as summary statistics of the data generating process, in the form of collections of intervals. This information can alternatively be represented by a {\em persistence diagram}, which takes the birth and death times of each bar in a barcode as an ordered pair $(x, y)$ and produces a scatterplot. This provides an alternative multi-scale topological summary of the shape or surface. In a persistence diagram, the points lie in $\mathbb{R}_{\geq 0}^2$ and all the points on the diagonal $x=y$ have infinite multiplicity. The diagonal is included to allow for well-defined metrics on the space of persistence diagrams (or barcodes). 

Since summary statistics are direct parallels to the invariants of a topological space, considering such topological approaches in data analysis is an intuitive way of reducing dimensionality in high-dimensional statistical problems (i.e.~where the number of predictors is far greater than the number of observations).




\clearpage
\newpage
\bibliography{SECT_Ref} 


\clearpage
\newpage
\section*{Figures and Tables}

\clearpage
\newpage

\begin{landscape}

\setlength{\extrarowheight}{3pt}
\begin{table}[H]
\centering
\caption{Detailed results for predicting disease free survival (DFS) and overall survival (OS) using Gaussian process regression models defined by the linear, Gaussian, and Cauchy covariance functions, respectively. For each model fit, we consider the predictive utility of four different genomic data types: gene expression, tumor morphometry, tumor geometry, and the proposed smooth Euler characteristic transform (SECT). Assessment is carried out by using the predictive squared correlation coefficient ($R^2$), where larger numbers indicate better performance. We also use Optimal\% to denote the percentage of the time that a model exhibits the greatest $R^2$. All values in bold represent the best method in these two assessment categories. These values are based on 1000 random 80-20 splits for each clinical outcome. Standard errors for each model are given the parentheses. Lastly, we give estimates for the bandwidth or length-scale parameter $\wh\theta$ used to compute each kernel function. Note that $\wh\theta$ was found by using 10-fold cross-validation over the grid $[0.1,10]$ with step sizes equal to 0.1.}
\begin{tabular}{|c|c|ccc|ccc|}
\hline
& &\multicolumn{3}{c|}{Disease Free Survival (DFS)}&\multicolumn{3}{c|}{Overall Survival (OS)}\\[2pt] \hline
Covariance Function(s) & Data Type & $R^2$ & Optimal\% & $\wh{\theta}$ & $R^2$ & Optimal\% &$\wh{\theta}$\\[2pt]\hline
\multirow{4.5}{*}{Linear Kernel} &Gene Expression & 0.097 (0.013) & 18.1\% & --- & 0.075 (0.01) & 18.4\% & --- \\[2pt]
&Morphometrics & 0.133 (0.015) & 23.6\% & --- & \textbf{0.127 (0.015)} & \textbf{33.9\%} & --- \\[2pt]
&Geometrics & 0.137 (0.017) & 23.9\% & --- & 0.100 (0.012) & 23.2\% & --- \\[2pt]
&SECT & \textbf{0.198 (0.023)} & \textbf{34.4\%} & --- & 0.097 (0.012) & 24.5\% & --- \\[2pt]\hline
\multirow{4.5}{*}{Gaussian Kernel}&Gene Expression & 0.129 (0.015) & 23.6\% & 4.3 & 0.082 (0.011) & 16.2\% & 10 \\[2pt]
&Morphometrics & 0.113 (0.013) & 16.3\% & 0.1 & 0.113 (0.012) & 22.3\% & 4.0 \\[2pt]
&Geometrics & 0.154 (0.018) & 21.1\% & 5.2 & 0.102 (0.013) & 22.3\% & 5.0 \\[2pt]
&SECT & \textbf{0.228 (0.026)} & \textbf{36.1\%} & 0.6 & \textbf{0.168 (0.018)} & \textbf{39.2\%} & 4.2 \\[2pt]\hline
\multirow{4.5}{*}{Cauchy Kernel}&Gene Expression & 0.126 (0.015) & 26.2\% & 6.4 & 0.084 (0.010) & 16.7\% & 10.0 \\[2pt]
&Morphometrics & 0.088 (0.012) & 15.9\% & 1.2 & 0.115 (0.014) & 27.9\% & 4.5 \\[2pt]
&Geometrics & 0.116 (0.013) & 17.4\% & 0.2 & 0.095 (0.012) & 18.9\% & 3.5 \\[2pt]
&SECT & \textbf{0.237 (0.027)} & \textbf{40.5\%} & 0.6 & \textbf{0.158 (0.017)} & \textbf{36.5\%} & 5.5 \\[2pt]\hline
\end{tabular}
\label{Tab1}
\end{table}

\end{landscape}


\setcounter{figure}{0}
\setcounter{table}{0}

\makeatletter 
\renewcommand{\thefigure}{S\@arabic\c@figure} 
\makeatletter 
\renewcommand{\thetable}{S\@arabic\c@table} 

\section*{Supplementary Figures}\addcontentsline{toc}{section}{Supplementary Figures}

\begin{figure}[H]
\centering
\includegraphics[width = 0.8\textwidth]{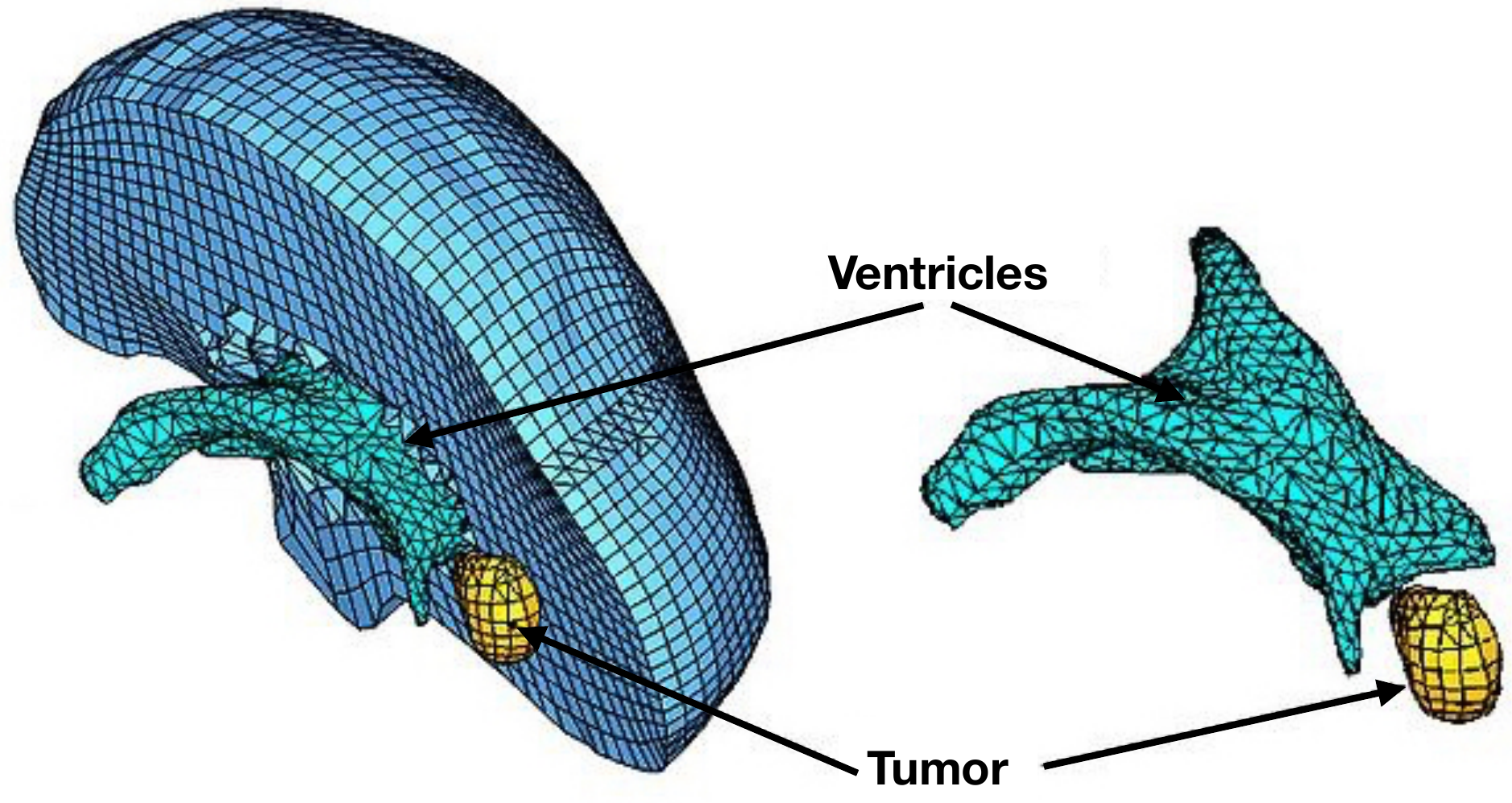}
\caption{A mesh representation of a brain tumor and ventricles.}
\label{Fig_S1}
\end{figure}

\begin{figure}[H]
\centering
\subfigure{
\includegraphics[width = 0.45\textwidth]{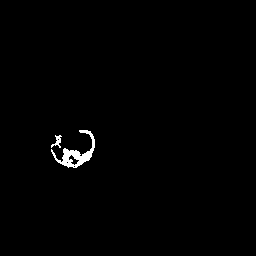}
\label{Fig_S2A}
}
\subfigure{
\includegraphics[width = 0.45\textwidth]{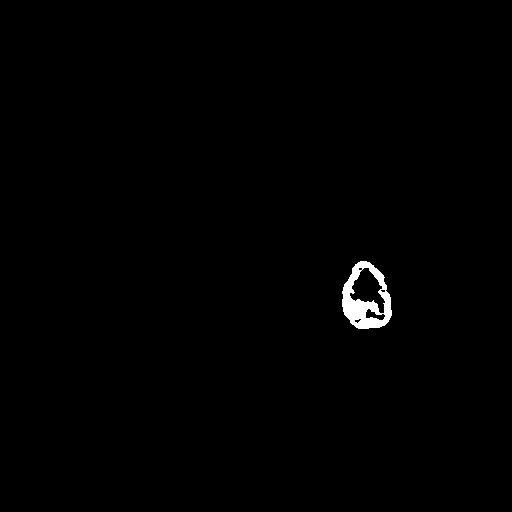}
\label{Fig_S2B}
}
\subfigure{
\includegraphics[width = 0.45\textwidth]{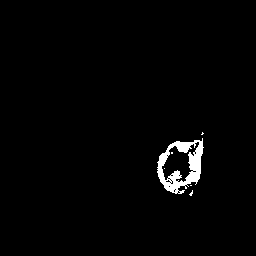}
\label{Fig_S2C}
}
\subfigure{
\includegraphics[width = 0.45\textwidth]{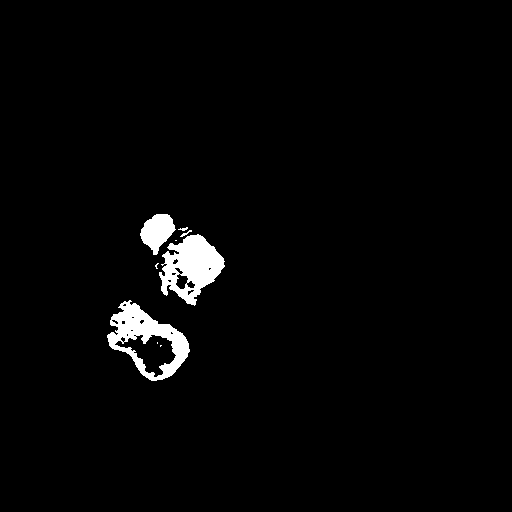}
\label{Fig_S2D}
}
\caption{Examples of tumors exhibiting necrosis and multifocality. Note that all four images are taken from different patients to highlight the diversity of disease progression. These images were segmented from the original MRI scans using the MITKats algorithm \cite{mitkats}.}
\label{Fig_S2}
\end{figure}

\begin{figure}[H]
\centering
\includegraphics[width=\textwidth]{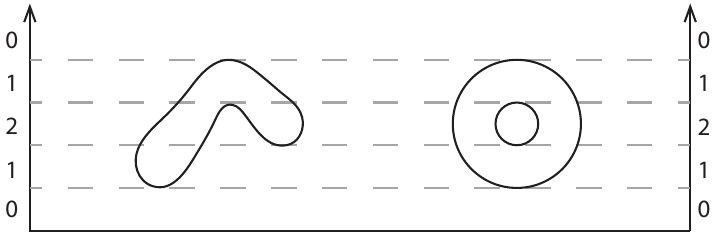}
\caption{Counterexample for injectivity of the Euler Characteristic (EC) curve for a fixed direction. The vertical axes show the direction of the filtration for both shapes by height (i.e.~the sublevel set filtration). The numbers on the axes denote the evolution of the EC for both shapes. We see that although the shapes are different, the corresponding ECs change in exactly the same manner, yielding identical ECTs for a fixed direction $\nu \in S^1$.}
\label{Fig_S3}
\end{figure}

\begin{figure}[H]
\centering
\subfigure[]{
\includegraphics[width = 0.45\textwidth]{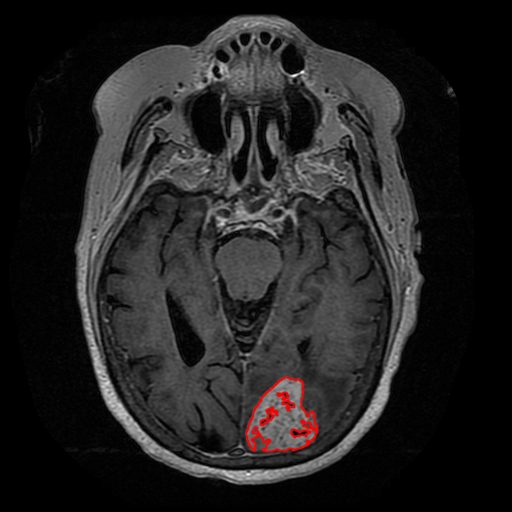}
\label{Fig_S4A}
}
\subfigure[]{
\includegraphics[width = 0.45\textwidth]{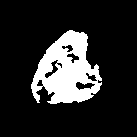}
\label{Fig_S4B}
}
\caption{Example image data used in radiomic analysis. An original MRI from the TCIA and TCGA is displayed in Figure (a), while the final segmented image via the MITKats algorithm \cite{mitkats} is given in Figure (b).}
\label{Fig_S4}
\end{figure}

\begin{figure}[H]
\centering
\includegraphics[scale=0.7]{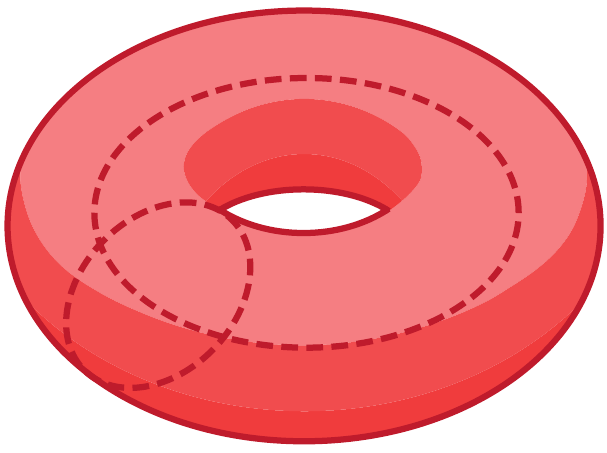}
\caption{An illustrative example of homology using the 2-dimensional torus and its cycles. The torus has a single connected component and a single $2$-cycle (the void locked inside the torus). In addition, it has two distinct $1$-dimensional cycles (or closed loops) represented by the two curves in the figure. Consequently, the Betti numbers of the torus are $\beta_0 = 1$, $\beta_1 = 2$, and $\beta_2 =1$.}
\label{Fig_S5}
\end{figure}

\begin{figure}[H]
\centering
\subfigure[]{
\includegraphics[width = 0.45\textwidth]{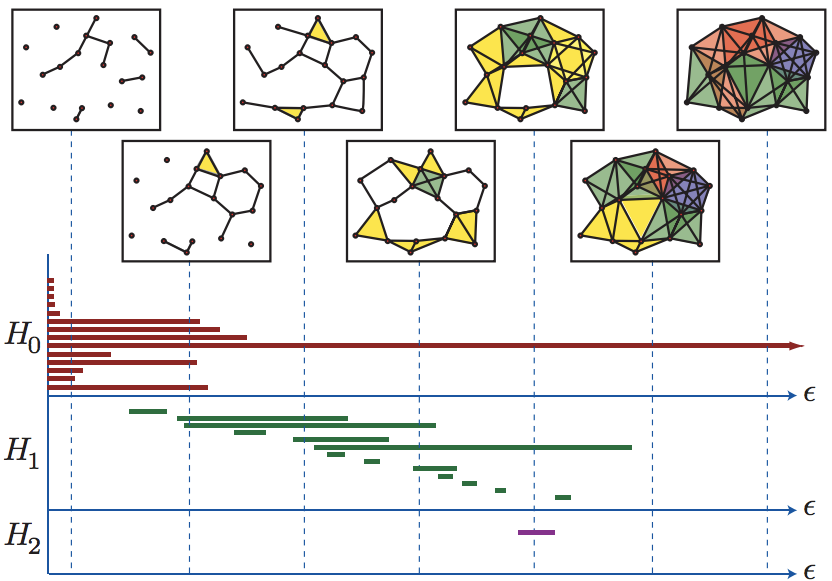}
\label{Fig_S6A}
}
\subfigure[]{
\includegraphics[width = 0.45\textwidth]{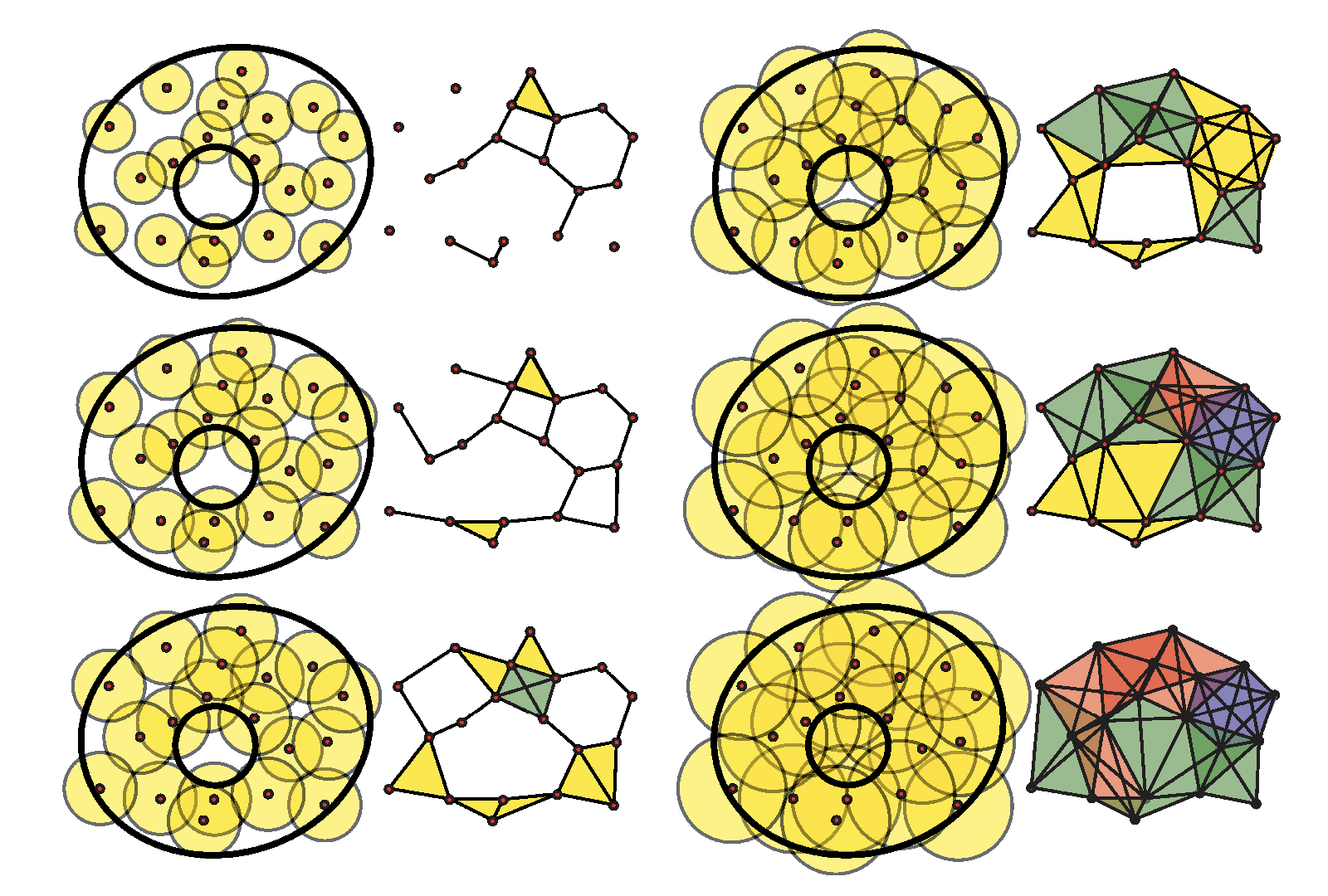}
\label{Fig_S6B}
}
\caption{Illustrative example of persistent homology and the resulting barcode. In Figure (a), the shape of interest is the annulus, whose topology is given by $\beta_0 = 1$, $\beta_1 = 1$, and $\beta_2 = 0$. Seventeen points are sampled from the annulus. Simplicial complexes are computed for continuous values of a filtration parameter, $\epsilon \in [0, \infty)$. Here, the filtration is given by the radii of balls centered at each sample point: the radius value is $\epsilon$, and the associated simplicial complex is built by replacing two overlapping balls by an edge, three overlapping balls by a face, and so on, for higher dimensions. The filtration, in terms of the evolving simplicial complex, is illustrated in the upper panels. Vertices, edges, and faces are formed as the value of $\epsilon$ increases. $H_0$ corresponds to connected components; $H_1$ corresponds to cycles whose boundaries are formed by edges between vertices; and $H_2$ corresponds to cycles with boundaries formed by faces. The dashed lines extending from the panels connect to the bars representing the topological features appearing and existing at the corresponding values of $\epsilon$. As $\epsilon$ progresses, connected components merge, cycles form, and fill up; the convention when two features merge is to retain the bar corresponding to the feature existing first. The barcode summarizes this progression of $\epsilon$ by tracking the ``lifetimes'' of the topological features according to their homology groups. Notice that there is a single $H_0$ bar that persists as $\epsilon \rightarrow \infty$, which represents the single connected component of the annulus. There are also several bars of varying length in $H_1$, including dominant bars, suggesting that the point cloud was unevenly sampled from the annulus in such a way that the sample space contains holes. Also, there is a single $H_2$ bar, which represents the cycle bounded by faces in the corresponding panel, but the length of the bar is comparatively short, and thus likely to be a spurious topological artifact. Figure (b) on the right shows the annulus with sampled points; the balls centered at the sampled points as the radius $\epsilon$ increases; and the corresponding simplicial complexes formed as the balls intersect (which correspond to the panels extending from specific bars in the barcode in Figure (a) on the left). This figure has been previously published \cite{Ghrist2008}.}
\label{Fig_S6}
\end{figure}


\end{document}